\documentclass[prb,twocolumn,notitlepage,superscriptaddress]{revtex4-1}
\usepackage{amsmath,amsfonts,amssymb,amsthm,epsfig,array}
\usepackage{dsfont}
\usepackage{graphics}
\usepackage{float}
\usepackage{verbatim}
\usepackage[dvipsnames]{xcolor}
\usepackage[hidelinks,colorlinks,linkcolor=blue,
citecolor=blue,urlcolor=blue]{hyperref}
\usepackage[titletoc,title]{appendix}
\usepackage{multirow}

\setcitestyle{number,sort,open={(},close={)}}

\newcommand{\bsl}[1]{\boldsymbol{#1}}

\newcommand{\ii}{\mathrm{i}}

\newcommand{\bra}[1]{\langle #1|}
\newcommand{\ket}[1]{|#1 \rangle}

\newcommand{\Tr}{\mathop{\mathrm{Tr}}}

\newcommand{\eqnref}[1]{Eq.\,\eqref{#1}}
\newcommand{\figref}[1]{Fig.\,\ref{#1}}

\newcommand{\refcite}[1]{Ref.\,[\onlinecite{#1}]}

\newcommand{\mat}[1]{\left(\begin{matrix}#1\end{matrix}\right)}

\newcommand{\eq}[1]{\begin{equation} #1 \end{equation}}

\newcommand{\eqa}[1]{\begin{align}\begin{split} #1 \end{split}\end{align}}
\let\oldAA\AA
\renewcommand{\AA}{\text{\normalfont\oldAA}}

\newcommand{\ie}{{\emph{i.e.}}}
\newcommand{\eg}{{\emph{e.g.}}}
\newcommand{\TR}{\mathcal{T}}

\usepackage[mathscr]{euscript}

\newcommand{\eqmxgyCN}{\eqnref{eq:mx_gy_CN} }
\newcommand{\eqthetamx}{\eqnref{eq:theta_mx} }
\newcommand{\eqnugy}{\eqnref{eq:nu_gy} }
\newcommand{\eqPgymod}{\eqnref{eq:P_gy_mod} }
\newcommand{\eqPgymodins}{\eqnref{eq:P_gy_insulator} }

\newcommand{\figSGtwosix}{\figref{fig:SG26}}

\begin{document}
\title{Gapless Criterion for Crystals from Effective Axion Field}
\author{Jiabin Yu}
\email{jky5062@psu.edu}
\affiliation{Department of Physics, the Pennsylvania State University, University Park, PA, 16802}
\author{Zhi-Da Song}
\affiliation{Department of Physics, Princeton University Princeton, NJ
08544, USA}
\author{Chao-Xing Liu}
\affiliation{Department of Physics, the Pennsylvania State University, University Park, PA, 16802}
\begin{abstract}
Gapless criteria that can efficiently determine whether a crystal is gapless or not are particularly useful for identifying topological semimetals.
In this work, we propose a sufficient gapless criterion for three-dimensional non-interacting crystals, based on the simplified expressions for the bulk average value of the static axion field.
The brief logic is that two different simplified expressions give the same value in an insulator, and thus the gapless phase can be detected by the mismatch of them.
We demonstrate the effectiveness of the gapless criterion in the magnetic systems with space groups 26 and 13, where mirror, glide, and inversion symmetries provide the simplified expressions.
%
In particular, the gapless criterion can identify gapless phases that are missed by the symmetry-representation approach, as illustrated by space group 26.
Our proposal serves as a guiding principle for future discovery of topological semimetals.
\end{abstract}
\maketitle


\section{Introduction}
Complementary to the study on topological insulators~\cite{Hasan2010TI,Qi2010TITSC}, the last decade has witnessed intense research interests in three dimensional (3D) topological semimetals~\cite{Burkov2016TSM,Yan2017WSM,Bernevig2018TSM,Vishwanath2018RMPTSM,
Wan2011WSM,Weng2015WSMTaAs,Xu2015WSM,Ding2015WSMTaAs,
Balents2011TNLSM,Fang2012DSM,Liu2014DSM,Soluyanov2015TypeIIWSM,
Weng2016TP,Soluyanov2016TP,Bzdusek2016NodalChain,Bernevig2016UnconTSM}.
The characterization of various topological semimetals and the search for material candidates are central topics in this field.
Recent progress~\cite{Turner2012NTRInv,Hughes2011NTRInver,Bradlyn2017TQC,
Po2017SymIndi,Kruthoff2017TCI,Song2018TSM,Po2020SI,
Bernevig2020MagneticTQC} classified topological phases based on the symmetry representations at high-symmetry momenta, which made possible the prediction of thousands of topological semimetals~\cite{Bernevig2019TopoMat,Fang2019TopoMat,Wan2019TopoMat,Bernevig2020MTQCMat}.
%
The high efficiency of this symmetry-representation approach originates from the fact that it only requires the information of the ground state wavefunciton in a lower dimensional submanifold of 3D first Brillouin zone (1BZ), like high-symmetry points/lines/planes.
%
However, many gapless phases cannot be identified with the symmetry-representation approach, especially when the gapless points exist at generic momenta, such as the Weyl semimetal TaAs~\cite{Weng2015WSMTaAs,Xu2015WSM,Ding2015WSMTaAs}. 
%
%
Therefore, new efficient approaches that can capture those missing phases are of particular importance.

In this work, we exploit the effective axion field $\theta$ of 3D materials and propose an efficient gapless criterion that can detect and provide topological invariants for the gapless phases beyond the symmetry-representation approach.
The dimensionless $\theta$ appears in the electromagnetic response of 3D insulating crystals as~\cite{Qi2008TFT,Essin2009AI}
\eq{
\label{eq:EdotB}
\mathcal{L}_\theta=-\frac{e^2}{h c}  \frac{\theta}{2\pi} \bsl{E}\cdot \bsl{B}\ ,
}
where $e$, $h$, and $c$ are the elementary charge, Planck constant, and speed of light, respectively.
%
%
Based on symmetries, several simplified expressions~\cite{Wang2010TRIAxion,
Turner2012NTRInv,Hughes2011NTRInver,Varjas2015AI,Ahn2019C2T,Sun2020C4TAI} have been derived for evaluating the average value of static $\theta$ in the bulk of materials.
The key idea of this work is to check the consistency of these simplified expressions.
Intuitively, the mismatch of two expressions reflects the ill-defined $\theta$, indicating the possible existence of gapless points, and the difference of the two mismatching expressions then serves as the topological invariant for the gapless phase.
The gapless criterion is efficient since the simplified expressions only involve lower dimensional submanifolds of 1BZ.
We demonstrate the effectiveness of this gapless criterion in two magnetic systems: one with space group (SG) 26 and spin-orbit coupling (SOC), and the other with SG 13 but without SOC.
In the first example, we find that the gapless criterion can identify Weyl semimetal phases beyond the symmetry-representation approach, and the semimetal hosts a surface mode with a saddle-shaped energy dispersion.
The gapless criterion agrees with the symmetry-representation approach in the second example.

\section{Effective Axion Field}

Before discussing the gapless criterion, we review the static effective axion field in 3D crystals and its simplified expressions given by symmetries.
We focus on insulators with vanishing quantum anomalous Hall (QAH) conductivity (or equivalently with zero Chern number (CN) on any 2D closed submanifold of 1BZ), like time-reversal (TR) invariant insulators~\cite{Qi2008TFT,Wang2010TRIAxion}.
In these insulators, a globally continuous gauge~\cite{Brouder2007Wannier} is allowed for the cell-periodic parts of the occupied Bloch states, labeled as $\ket{u_{\bsl{k},a}}$ with $a=1,...,N$ and $N$ the number of occupied bands.
With this gauge, the bulk average value of the static effective axion field can be expressed as~\cite{Qi2008TFT,Essin2009AI,Wang2010TRIAxion,Turner2012NTRInv,Ahn2019C2T}
\eq{
\label{eq:theta_A}
\theta=\frac{1}{4\pi}\int d^3k\ \epsilon^{ijl}\Tr[A_i\partial_{k_j}A_l+\ii \frac{2}{3}A_i A_j A_l]\ ,
} 
where $\theta$ labels the bulk average value henceforth, duplicated indexes are summed over, and $[A_i(\bsl{k})]_{a_1a_2}=-\ii \bra{u_{\bsl{k},a_1}}\partial_{k_i}\ket{u_{\bsl{k},a_2}}$ is the Berry connection.
%
%
%
%
The gauge transformation allowed in \eqnref{eq:theta_A}, $\ket{u_{\bsl{k},a}}\rightarrow \ket{u_{\bsl{k},a'}}[U(\bsl{k})]_{a' a}$, can only change $\theta$ by multiples of $2\pi$, since the gauge choice requires $U(\bsl{k})$ to be globally continuous.
Thus, $\theta \mod 2\pi$ is unambiguous.
%
%
%
%
%
%
%
%
\eqnref{eq:theta_A} is hard to use in general, since we need to know the occupied Bloch wavefunctions in the entire 1BZ and derive the globally continuous bases from them.

In the presence of a so-called ``axion-odd" symmetry~\cite{Qi2008TFT,Essin2009AI,Hughes2011NTRInver,Turner2012NTRInv,
Fang2012TIPG,Varjas2015AI,Wang2010TRIAxion,
Varnava2018AI,Schindler2018HOTI,Wieder2018AXIFragile,
Ahn2019C2T,Varnava2019AIWannier,Sun2020C4TAI}, $\theta\mod 2\pi$ is quantized to $0$ or $\pi$.
An axion-odd symmetry is either an improper SG symmetry or a combination of TR symmetry and a proper SG symmetry, where ``proper" or ``improper" means that the point group part of the SG operation has determinant $1$ or $-1$, respectively, when acting on the real space position.
In appropriate setups, physical consequences of $\theta=\pi$ include quantized magnetoelectric effect~\cite{Qi2008TFT,Yu2019MRAI}, quantized zero Hall plateau~\cite{Wang2015AI,Mogi2017AI,Xiao2018AI}, and Faraday and Kerr rotation~\cite{Wu2016THzTI,Dziom2017THzTI,Okada2016THzQAH}.
Magnetic materials (MnTe$)_n($Bi$_2$Te$_3)_m$ were recently proposed as candidates for the $\theta=\pi$ phase~\cite{Zhang2018AIMnBi2Te4,Vishwanath2019MnBiTeAI}.

\begin{figure}[t]
    \centering
    \includegraphics[width=\columnwidth]{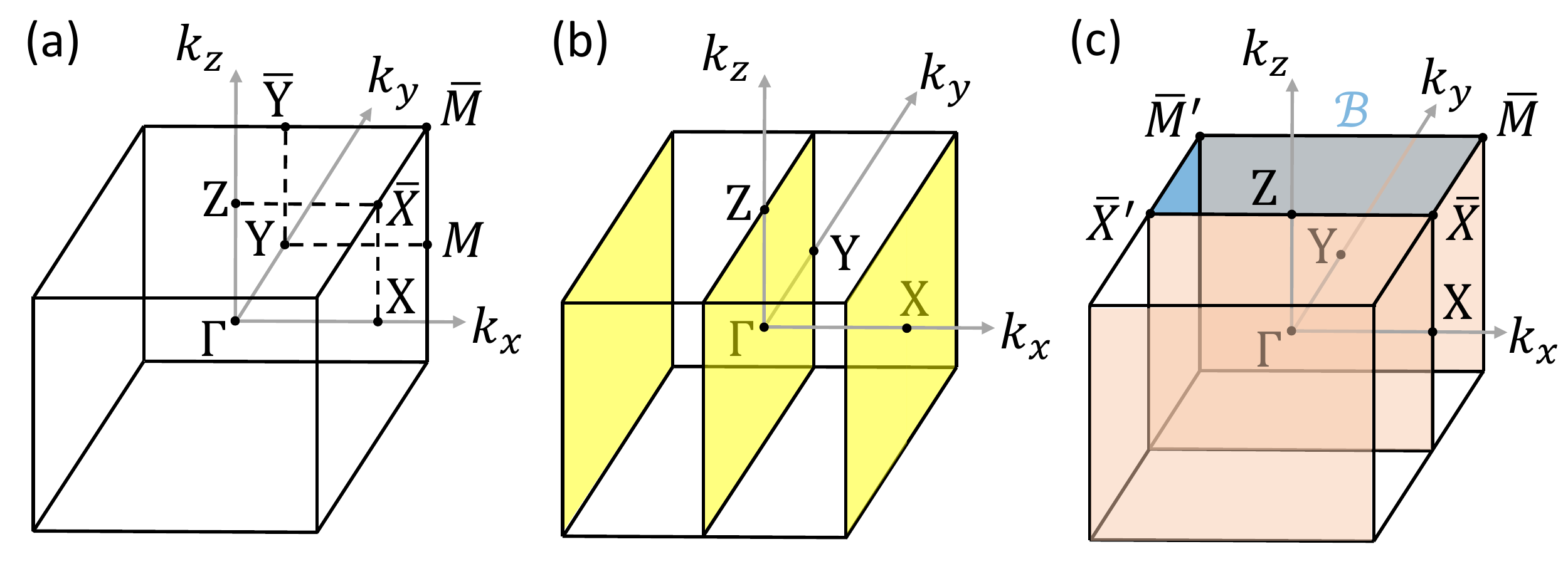}
    \caption{(a), (b) and (c) show the 1BZs for the inversion $P$, mirror $m_x$ and glide $g_y$ cases, respectively. 
    Without loss of generality, we, throughout the work, choose three primitive reciprocal lattice vectors to be orthogonal and set their magnitudes to be 1.
    In (a), the black dots label the eight inequivalent inversion-invariant momenta.
    The yellow planes in (b) are at $k_x=-\pi,0,\pi$ and invariant under $m_x$.
    %
    In (c), the glide-invariant planes at $k_y=-\pi,0,\pi$ are marked pink, and the blue area, labeled by $\mathscr{B}$, is parametrized as $k_x\in (-\pi,\pi]$, $k_y\in [0,\pi]$ and $k_z=\pi$.
    }
    \label{fig:AOS}
\end{figure}

We define an axion-odd symmetry $g$ to be an ``axion-odd-simplification" (AOS) symmetry if it can provide a gauge-invariant expression $\nu_g$ for $\theta$ that only involves a lower-dimensional submanifold of 1BZ.
$\nu_g$ dramatically simplifies the evaluation of $\theta$ since it does not involve the whole 1BZ or require the globally continuous gauge.
%
%
A widely known AOS symmetry is the inversion symmetry $P$, whose simplified expression $\nu_P$ for $\theta$ reads~\cite{Turner2012NTRInv,Hughes2011NTRInver}
\eq{
\label{eq:nu_P}
\frac{\theta}{\pi}\mod 2 =\nu_P=\sum_{\bsl{K}} \frac{n_{\bsl{K}}^{P,+}-n_{\bsl{K}}^{P,-}}{4}\mod 2\ ,
}
where 
$n^{P,\pm}_{\bsl{K}}$ labels the number of occupied states with inversion eigenvalue $\pm 1$ at the inversion-invariant momentum $\bsl{K}$, and the sum of $\bsl{K}$ ranges over all inversion-invariant momenta as shown in \figref{fig:AOS}(a).
%

Mirror is also an AOS symmetry.
Without loss of generality, we consider the mirror operation $m_x$ that flips $x$.
The states with $k_x=\Lambda$ have definite mirror eigenvalues $\pm s$, where $\Lambda=0,\pi$ as shown in \figref{fig:AOS}(b), $s=1$ without SOC, and $s=\ii$ with SOC.
%
%
We can then define the mirror CN~\cite{Teo2008MCN} as 
$
C^M_{k_x=\Lambda}=(C^{m_x,+}_{k_x=\Lambda}-C^{m_x,-}_{k_x=\Lambda})/2
$,
where $C^{m_x,\pm}_{k_x=\Lambda}$ is the total CN of occupied bands with mirror eigenvalue $\pm s$.
Finally, the simplified expression $\nu_{m_x}$ for $\theta$ reads~\cite{Varjas2015AI,Varnava2019AIWannier}
\eq{
\label{eq:theta_mx}
\frac{\theta}{\pi}\mod 2= \nu_{m_x}= C_{k_x=0}^M-C_{k_x=\pi}^M\mod 2\ .
}

Similar situation happens for the glide symmetry.
Without loss of generality, we consider the combination of half lattice translation along $z$ and the mirror operation $m_y$ that flips $y$, labeled as $g_y=\left\{m_y|00\frac{1}{2}\right\}$.
The states with $k_y=\Lambda$ can have definite glide eigenvalues $\pm s e^{-\ii k_z/2}$.
Then, the simplified expression $\nu_{g_y}$ for $\theta$ reads~\cite{Fang2015TCI,Ken2015Mobius,Wang2015AI,Kim2019Glide,
Varnava2019AIWannier}
\eqa{
\label{eq:nu_gy}
&\frac{\theta}{\pi}\mod 2=\nu_{g_y}=\\
&C^{g_y,-}_{k_y=0}-C^{g_y,-}_{k_y=\pi}+C_{\mathscr{B}} -2(\gamma^{g_y,-}_{\bar{X}'\bar{X}}+\gamma^{g_y,-}_{\bar{M}\bar{M}'})\mod 2\ .
}
$C_{k_y=\Lambda}^{g_y,-}$ and $C_{\mathscr{B}}$ are integrals of Berry curvature divided by $2\pi$ for the occupied bands with glide eigenvalue $-s e^{-\ii k_z/2}$ on plane $k_y=\Lambda$ and for all occupied bands in the area $\mathscr{B}$, respectively. (See \figref{fig:AOS}(c) for details.)
$\gamma^{g_y,-}_{\bar{X}'\bar{X}/\bar{M}\bar{M}'}$ labels the Berry phase divided by $2\pi$ for the occupied bands with glide eigenvalue $-s e^{-\ii k_z/2}$ along path $\bar{X}'\bar{X}/\bar{M}\bar{M}'$. 

Other known AOS symmetries include TR~\cite{Wang2010TRIAxion} and the combination of TR and n-fold ($n=2,4$) rotational symmetries~\cite{Ahn2019C2T,Sun2020C4TAI} in the presence of SOC.
All the examples show that the submanifold, on which the simplified expression is defined, mostly consists of high-symmetry momenta.
Up to now, it is still unclear whether all axion-odd symmetries are AOS symmetries.

\section{Gapless Criterion}
The gapless criterion that we propose is for crystals with at least two AOS symmetries.
Based on the currently confirmed AOS symmetries discussed above, 
the condition is satisfied by 217 SGs in the presence of TR symmetry and SOC and by 155 SGs otherwise~\cite{SMs}, which is quite common.
Let us generically pick two AOS symmetries in the crystal, labeled as $g_1$ and $g_2$.
If the crystal is gapped and has zero CNs, the two simplified expressions, $\nu_{g_1}$ and $\nu_{g_2}$, given by $g_1$ and $g_2$ for $\theta$ must be well-defined and equal (both equal to $\theta/\pi \mod 2$).
The contrapositive of the above statement is that the crystal is either gapped with non-zero CNs or gapless if (i) at least one of the two simplified expressions is ill-defined, or (ii) both simplified expressions are well defined but mismatch $(\nu_{g_1}\neq \nu_{g_2})$.
%
Therefore, if the condition of the contrapositive is satisfied, the system must be in a non-trivial phase.
In the case (i), the crystal must be gapless, since a simplified expression becomes ill-defined only if gapless points exist in the corresponding submanifold of 1BZ. 
However, the gapless points in this case are relatively simple to locate as the submanifold mostly consists of high symmetry momenta.
So we focus on the case (ii) where gapless points stay away from the submanifolds.

In the case (ii), we need to rule out the possibility of insulators with non-zero CNs (also called QAH insulators) to get gepless phases, which can be achieved by certain symmetries, like the TR symmetry.
In the absence of TR symmetry, the symmetries that forbid QAH insulators exist in 141 out of the 155 SGs that contain at least two AOS symmetries~\cite{SMs}, like SG 26 as discussed later.
%
%
It means that if the case (ii) happens in the presence of TR symmetry or for any of the 141 SGs, we directly know that the crystal is gapless. 
%
%
%
%
The remaining 14 out of 155 SGs cannot directly forbid QAH insulators based on symmetries in the absence of TR symmetry, like SG 13 as discussed later.
For them, we can rule out the QAH insulators by checking the following ``Existence-of-Zero-CN" (EZCN) condition:
%
parallel to \emph{every} two of the three primitive reciprocal lattice vectors, there \emph{exist} a 2D gapped plane in 1BZ on which the system has zero CN.
Here a gapped plane means that the system is gapped everywhere on the plane.
%
The EZCN condition is equivalent to vanishing QAH conductivity when the crystal is insulating, but it is satisfiable in gapless crystals.
%
The process of checking the condition can be simplified by symmetries, \eg, the CN on any plane that is perpendicular to a mirror/glide plane must be zero.

\begin{figure}[t]
    \centering
    \includegraphics[width=0.9\columnwidth]{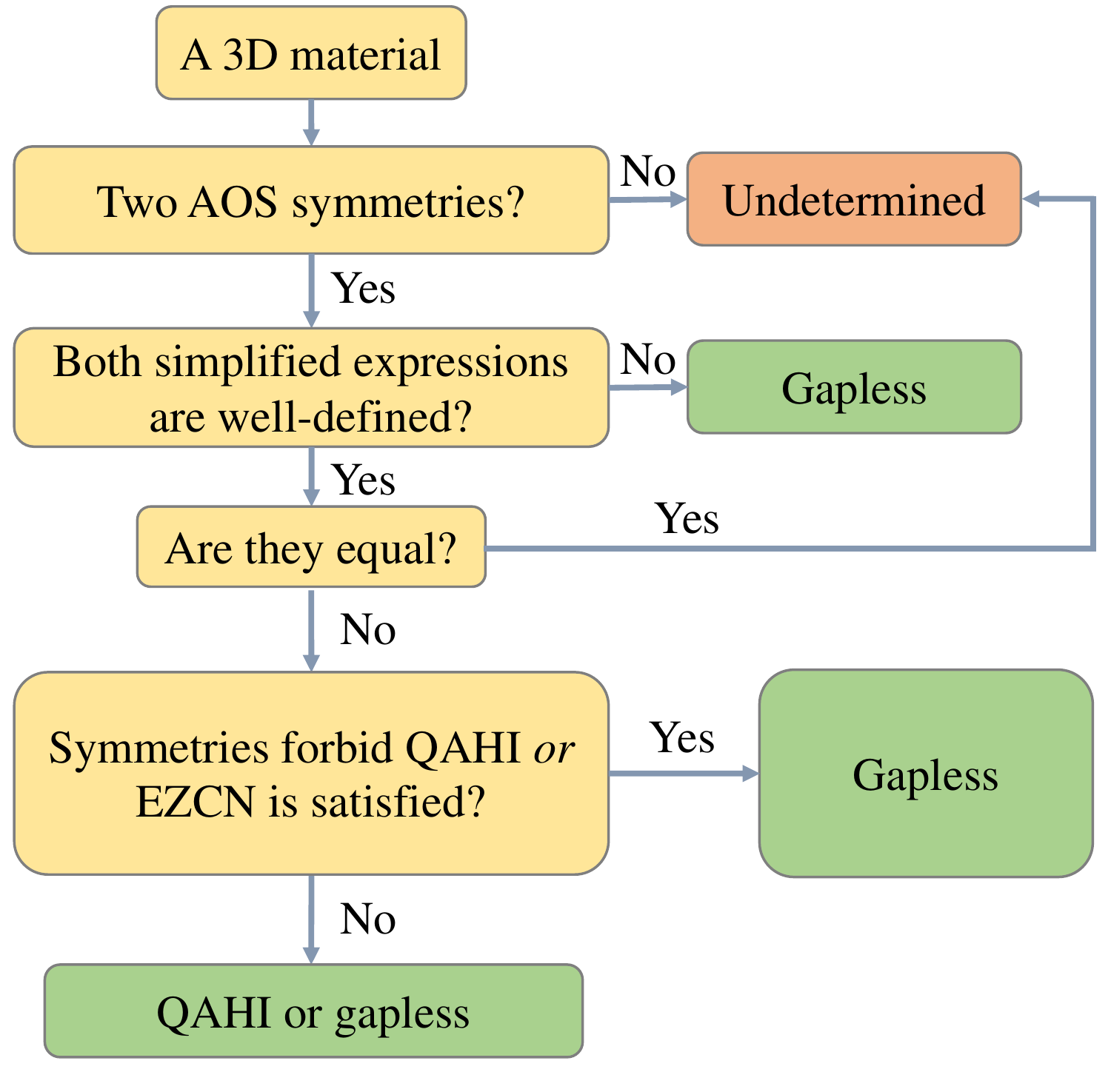}
    \caption{
    This flowchart demonstrates how to use the gapless criterion given a 3D material.
    Here ``QAHI" means the QAH insulator phase.
    }
    \label{fig:flowchart}
\end{figure}

\figref{fig:flowchart} summarizes the above logic in a flowchart, from which we derive a gapless criterion: given a crystal that has at least two AOS symmetries and cannot be a QAH insulator (indicated by either symmetries or by explicitly satisfying the EZCN condition), it is gapless if there exist two well-defined simplified expressions that mismatch.
This is the main result of this work.
Under the protection of the condition that forbids QAH insulators, the gapless phase identified by this gapless criterion is stable against any perturbations that preserve the nonzero $\nu_{g_1}-\nu_{g_2}$, making it a topological invariant.
In the following, we demonstrate the effectiveness of the criterion for SG 26 and SG 13.

\section{Mangetic Systems with SG 26}

\begin{figure}[t]
    \centering
    \includegraphics[width=\columnwidth]{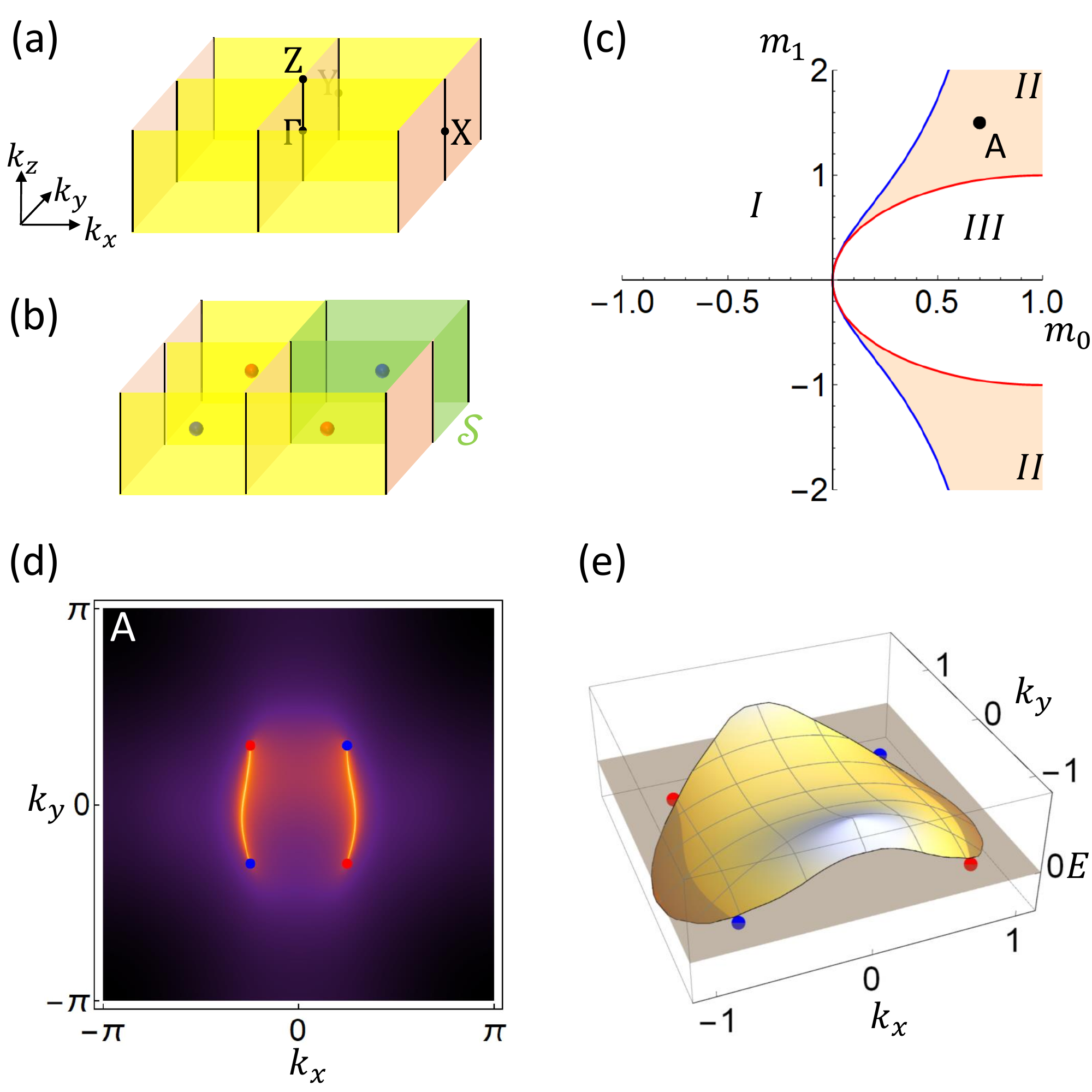}
    \caption{
    (a) shows the 1BZ of SG 26. 
    The yellow and pink planes are invariant under $m_x$ and $g_y$ symmetries, respectively.
    The black lines are the intersections of those planes, where the little group is the entire SG.
    (b) demonstrates a distribution of Weyl points (red or blue dots) in a possible Weyl semimetal phase.
    The dots of the same (different) colors have the same (opposite) chiralities.
    The green area $\mathscr{S}$ surrounds the $k_{x,y}>0$ quarter of 1BZ. 
    (c) is the phase diagram generated by the toy model. 
    Phase I and III are insulating phases with $\theta=0$ and $\theta=\pi$, respectively.
    Phase II is a Weyl semimetal phase that has the distribution of Weyl points as (b).
    %
    %
    %
    (d) shows the (001) surface density of states at zero energy and (e) demonstrates the saddle-shaped energy dispersion of the (001) surface gapless modes.
    Both graphs are plotted for point A in (c), and the gray plane in (e) is at zero energy.
    %
    %
    }
    \label{fig:SG26}
\end{figure}

We first study the spin-orbit coupled magnetic materials whose magnetic SG is the same as SG 26 ($Pmc2_1$).
Besides lattice translations, SG 26 is generated by two AOS symmetries $m_x$ and $g_y$.
%
%
The inequivalent high symmetry momenta include two mirror-invariant planes ($k_x=0,\pi$), two glide-invariant planes ($k_y=0,\pi$), and their intersections, as shown in \figref{fig:SG26}(a).
We consider the case where the crystal is gapped at all these high-symmetry points.
%
%
%
In this case, the symmetry representations furnished by the occupied bands at high-symmetry momenta are always the same as those of atomic insulators, according to \refcite{Po2017SymIndi,Bradlyn2017TQC}.
However, the crystal can still be gapless, for example, having 4 Weyl points at generic momenta as shown in \figref{fig:SG26}(b).
Therefore, such gapless phase cannot be identified by the symmetry-representation approach~\cite{SMs}.

On the contrary, our gapless criterion is effective.
The gapped high-symmetry points make both $\nu_{m_x}$ and $\nu_{g_y}$ well-defined according to \eqnref{eq:theta_mx} and \eqnref{eq:nu_gy}, since $C_{\mathscr{B}}$ is identically zero owing to $m_x$.
$m_x$ and $g_y$ further require the crystal to have zero CN on any gapped plane that is perpendicular to any $k$ axis, ruling out the possibility of QAH insulators.
Then, the gapless criterion indicates that the system is gapless when $\nu_{m_x}\neq \nu_{g_y}$.
%
%
%
%
%
Further derivation shows
\eq{
\label{eq:mx_gy_CN}
\nu_{m_x}+\nu_{g_y}\mod 2=C_{\mathscr{S}}\mod 2\ ,
}
where $C_{\mathscr{S}}$ is the so-called ``bent CN"~\cite{Bernevig2014BendCN,SMs} over area $\mathscr{S}$ in \figref{fig:SG26}(b).
When $\nu_{m_x}\neq \nu_{g_y}$, $C_{\mathscr{S}}$ must be an odd number, and thus the corresponding gapless phase must contain an odd number of Weyl points in the quarter of 1BZ surrounded by $\mathscr{S}$, \eg, \figref{fig:SG26}(b).

To verify the above analysis, we construct a toy model~\cite{SMs} by putting a spinful $s$-orbital at the origin and symmetrizing it with SG 26.
The resulting Hamiltonian has two tuning parameters $m_0$ and $m_1$, and we map out the phase diagram at half filling in \figref{fig:SG26}(c).
There are two insulating phases, phase I with $(\nu_{m_x},\nu_{g_y})=(0,0)$ and phase III with $(\nu_{m_x},\nu_{g_y})=(1,1)$.
As tuning the system from phase I to phase III, the gap closes on $k_y=0$ plane at the blue boundary and $\nu_{g_y}$ changes from 0 to 1, resulting in phase II with $(\nu_{m_x},\nu_{g_y})=(0,1)$.
Phase II further evolves into phase III across the red boundary, where the gap closes on $k_x=0$ plane and $\nu_{m_x}$ changes to 1.
According to the gapless criterion and \eqnref{eq:mx_gy_CN}, phase II should be a WSM phase with odd $C_{\mathscr{S}}$.
Indeed, the phase contains 4 Weyl points symmetrically distributed on the $k_z=0$ plane just like \figref{fig:SG26}(b), and we show the Fermi arcs on (001) surface as the bright arcs in \figref{fig:SG26}(d).
%
%
The Fermi arcs are parts of the saddle-shape energy dispersion~\cite{Fu2013SurfaceSaddlePoints,Singh2018SurfaceSaddlePoints} on (001) surface, as shown in \figref{fig:SG26}(e). 
Away from the Weyl points, the surface mode typically lies in the direct bulk gap~\cite{SMs}, which can be detected in ARPES experiments.
%

\section{Magnetic Crystals with SG 13}

\begin{figure}[t]
    \centering
    \includegraphics[width=\columnwidth]{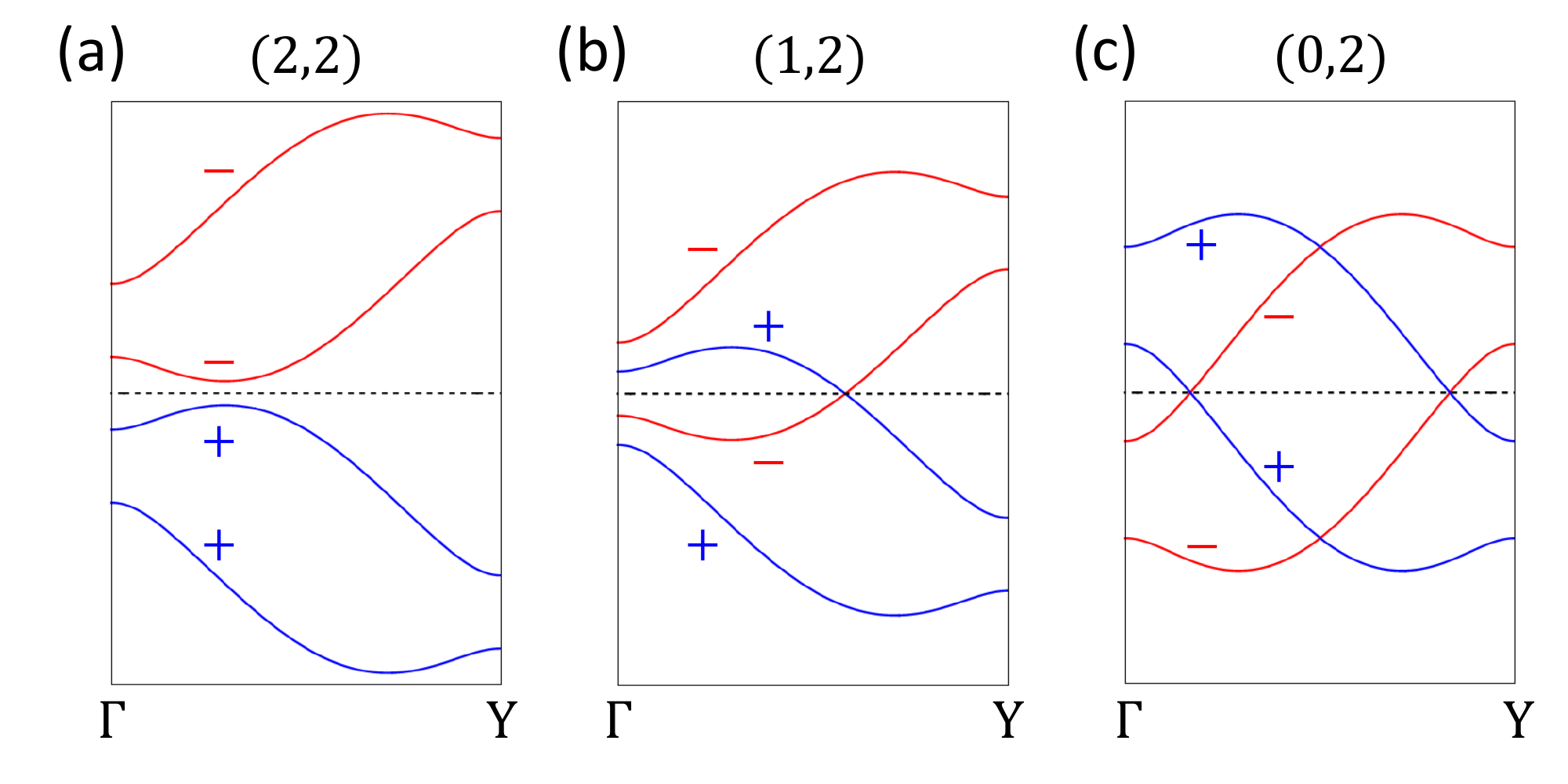}
    \caption{This figure includes schematic plots of band structures for SG 13 along $\Gamma-Y$.
    Blue and red lines are bands with $R_2$ eigenvalue $+1$ and $-1$, respectively, and the black dotted line is the Fermi energy.
    The numbers in the brackets are values of $(n^{\Gamma}_{R_2,+},n^{Y}_{R_2,+})$.
    }
    \label{fig:SG13}
\end{figure}

In this section, we study the magnetic crystal whose magnetic SG is SG 13 ($P2/c$) and whose SOC is neglectible.
SG 13 contains two AOS symmetries, inversion $P$ and glide $g_y$.
%
Since the combined $R_2=g_y P$ symmetry makes $C_{\mathscr{B}}$ identically zero, we choose the crystal to be gapped on the glide-invariant planes ($k_y=0,\pi$) to keep $\nu_{g_y}$ and $\nu_P$ well-defined. (See \figref{fig:AOS}(a,c).)
As the symmetries cannot forbid the QAH insulators, we further set the CN on $k_y=\pi$ to zero and keep the crystal gapped on $k_x=\pi$ and $k_z=\pi$ planes to satisfy the EZCN condition.
%
%
The CNs on $k_x=\pi$ and $k_z=\pi$ planes naturally vanish owing to $g_y$.

With these conditions, the gapless criterion indicates the crystal is gapless if $\nu_P - \nu_{g_y}\neq 0$.
To understand its physical meaning, we first note that any state with $(k_x,k_z)=(0,0)$ can have definite $R_2$ eigenvalues $\pm 1$.
Then, by generalizing the result in \refcite{Kim2019Glide}, we obtain~\cite{SMs}
\eq{
\label{eq:P_gy_mod}
\nu_P = \nu_{g_y} +\frac{n^{\Gamma}_{R_2,+}-n^{Y}_{R_2,+}}{2} \mod 2\ ,
}
where $n^{\Gamma/Y}_{R_2,+}$ is the number of occupied states with $R_2$ eigenvalue $+1$ at $\Gamma/Y$.
%
When the crystal is insulating, $n^{\Gamma}_{R_2,+}=n^{Y}_{R_2,+}$ must hold as exemplified in \figref{fig:SG13}(a), consistent with $\nu_P = \nu_{g_y}$.
When $\nu_P \neq \nu_{g_y}$, the nonzero $(n^{\Gamma}_{R_2,+}-n^{Y}_{R_2,+})$ indicates the existence of $R_2$-protected gapless points along $\Gamma-Y$ according to the symmetry-representation approach, as shown in \figref{fig:SG13}(b-c), and the gapless points are Weyl points without any fine tuning.
%
%
Therefore, the gapless criterion $\nu_P\neq \nu_{g_y}$ can identify the gapless phases detectable by the symmetry-representation approach.

In the above discussion, we do not consider the TR symmetry and explicitly impose the EZCN condition.
If we impose TR symmetry (making the crystal nonmagnetic), the gapless phases identified by $\nu_P \neq \nu_{g_y}$ would become nodal-line semimetal phases owing to the combination of the TR and inversion symmetries~\cite{Tomas2019NonAbelianNodalLine}.
%
If we abandon the EZCN condition and consider a generic TR-breaking insulator with SG 13 and neglectible SOC, we would have~\cite{SMs}
\eq{
\label{eq:P_gy_insulator}
\nu_P=\nu_{g_y} + C_{k_y=0} \mod 2\ ,
}
where $C_{k_y=0}$ is the CN on the glide-invariant plane, and thus $\nu_P \neq \nu_{g_y}$ suggests the insulator has odd $C_{k_y=0}$.

\section{Conclusion and Discussion}
Based on the effective axion field, we propose a gapless criterion for 3D crystals that have at least two AOS symmetries.
The criterion is potentially applicable to systems with or without SOC and can identify gapless phases beyond the symmetry-representation approach.
When applying this criterion in practice (say in first-principle calculations), it is better to follow \figref{fig:flowchart} and check the possibility of the QAH insulators at the very end, since the mismatch of two simplified expressions already indicates a non-trivial phase, QAH insulator or gapless.
%

As the gapless criterion can be more powerful if more AOS symmetries are identified, our proposal provides a driving force for future related theoretical study.
Recently, \refcite{Fang2019GaplessS4+T} studied the spin-orbit coupled systems with TR and $S_4$ (four-fold rotation combined with inversion) symmetries, and demonstrated that some of its Weyl semimetal phases can be detected by the mismatch of two $Z_2$ indexes that are respectively protected by the two symmetries.
Therefore, it is intriguing to ask whether the $Z_2$ index of axion-odd $S_4$ is a simplified expression of $\theta$.
%
%
Moreover, the non-integer value of one certain simplified expression (like $\nu_P$ or $\nu_{m_x}$) can directly detect the gapless phase, which is also worth exploring in the future.

\section{Acknowledgement}

J.Y. thanks Biao Lian, Hoi Chun Po, and Xiao-Qi Sun for helpful discussions.
Z.D.S. is supported by the U.S.
Department of Energy (grant no. DE-SC0016239), the National Science
Foundation (EAGER grant no. DMR 1643312), Simons Investigator
awards (grant no. 404513), ONR (grant no. N00014-14-1-0330),
NSF-MRSEC (grant no. DMR-142051), the Packard Foundation, the
Schmidt Fund for Innovative Research, and the Guggenheim Fellowship.
J.Y. and C.X.L. acknowledge the support of the Office of Naval Research (Grant No. N00014-18-1-2793), the U.S. Department of Energy (Grant No.~DESC0019064), and Kaufman New Initiative research grant KA2018-98553 of the Pittsburgh Foundation.

\begin{widetext}
\appendix

\section{More Details on Counting Applicable SGs}
In this section, we show more details on counting SGs for which our gapless criterion is potentially applicable.
The counting here is only for the 230 SGs, instead of the whole set of magnetic SGs.
Before moving onto the detailed discussion, let us discuss a general feature.
Given an AOS symmetry $g$, $g T_{\bsl{R}}$ with $T_{\bsl{R}}$ a translation by a lattice vector $\bsl{R}$ is also an AOS symmetry.
However, we should not treat $g$ and $g T_{\bsl{R}}$ as two different AOS symmetries.
It is because $T_{\bsl{R}}$ just adds a $e^{-\ii \bsl{R}\cdot\bsl{k}}$ phase to any eigenvalue of $g$ without changing the corresponding eigenvector.
For example, $P\ket{\psi_{\bsl{K},\pm}}=\pm \ket{\psi_{\bsl{K},\pm}}$ is equivalent to $P  T_{\bsl{R}}\ket{\psi_{\bsl{K},\pm}}=\pm e^{-\ii \bsl{R}\cdot\bsl{K}}\ket{\psi_{\bsl{K},\pm}}$, where $P$ is inversion and $\bsl{K}$ is an inversion-invariant momentum.
Then, the simplified expressions given by $g$ and $g T_{\bsl{R}}$ are equivalent and always have the same values; the only difference is the meaning of the eigenvalue labels in the expressions.
Again in the example of $P$ and $P  T_{\bsl{R}}$, $n^{P,\pm}_{\bsl{K}}$ labels the number of occupied states with $P$ eigenvalue $\pm 1$ at $\bsl{K}$, while $n^{P T_{\bsl{R}},\pm}_{\bsl{K}}$ corresponds to $P T_{\bsl{R}}$ eigenvalue $\pm e^{-\ii \bsl{R}\cdot\bsl{K}}$ at $\bsl{K}$.
Nevertheless, we always have $n^{P,\pm}_{\bsl{K}}=n^{P T_{\bsl{R}},\pm}_{\bsl{K}}$.

\subsection{With TR Symmetry and SOC}

We first consider the case where there is TR symmetry and SOC.
In the presence of SOC, the currently confirmed AOS symmetries include TR $\TR$, $C_2 \TR$, $C_4 \TR$, inversion, mirror, and glide symmetries, where $C_n$ means n-fold rotation. 
Therefore, in the presence of TR symmetry and SOC, any SG that contain $C_2$, $C_4$, inversion, mirror, or glide symmetry is a SG that contains at least two AOS symmetries.

To have a list of all these SGs, let us first separately list all SGs for each relevant symmetry.
Any SG that contains $C_2$ must contain SG 3 as a subgroup, and then we can use the subgroup relations of the SGs to get 
\eqa{
&\text{The list of all SGs that contain $C_2$}= \\
&\{3,5,10,12,13,15,16,17,18,20,21,22,23,24,25,27,28,30,32,34,35,37,38,39,40,41,42,43,44,45,46,47,48,49,\\
&50,51,52,53,54,55,56,57,58,59,60,63,64,65,66,67,68,69,70,71,72,73,74,75,77,79,80,81,82,83,84,85,86,87,\\
&88,89,90,91,92,93,94,95,96,97,98,99,100,101,102,103,104,105,106,107,108,109,110,111,112,113,114,115,116,\\
&117,118,119,120,121,122,123,124,125,126,127,128,129,130,131,132,133,134,135,136,137,138,139,140,141,142,\\
&149,150,151,152,153,154,155,162,163,164,165,166,167,168,171,172,175,177,178,179,180,181,182,183,184,187,\\
&188,189,190,191,192,193,194,195,196,197,199,200,201,202,203,204,206,207,208,209,210,211,212,213,214,215,\\
&216,217,218,219,220,221,222,223,224,225,226,227,228,229,230\}\ .
} 
Any SG that contains $C_4$ must contain SG 75 as a subgroup, and then we can use the subgroup relations of the SGs to get 
\eqa{
&\text{The list of all SGs that contain $C_4$}= \\
&\{75,79,83,85,87,89,90,97,99,100,103,104,107,108,123,124,125,126,127,128,129,130,139,140,207,209,211,221,\\
&222,225,226,229\}\ .
}
Any SG that contains inversion must contain SG 2 as a subgroup, and then we can use the subgroup relations of the SGs to get 
\eqa{
\label{eq:inversion_set}
&\text{The list of all SGs that contain inversion}= \\
&\{2,10,11,12,13,14,15,47,48,49,50,51,52,53,54,55,56,57,58,59,60,61,62,63,64,65,66,67,68,69,70,71,72,73,74,\\
&83,84,85,86,87,88,123,124,125,126,127,128,129,130,131,132,133,134,135,136,137,138,139,140,141,142,147,148,\\
&162,163,164,165,166,167,175,176,191,192,193,194,200,201,202,203,204,205,206,221,222,223,224,225,226,227,228,\\
&229,230\}\ .
} 
Any SG that contains mirror(s) must contain SG 6 as a subgroup, and then we can use the subgroup relations of the SGs to get 
\eqa{
\label{eq:mirror_set}
&\text{The list of all SGs that contain mirror(s)}= \\
&\{6,8,10,11,12,25,26,28,31,35,36,38,39,40,42,44,46,47,49,51,53,55,57,58,59,62,63,64,65,66,67,69,71,72,74,\\
&83,84,87,99,100,101,102,105,107,108,109,111,113,115,119,121,123,124,125,127,128,129,131,132,134,135,136,\\
&137,138,139,140,141,156,157,160,162,164,166,174,175,176,183,185,186,187,188,189,190,191,192,193,194,200,\\
&202,204,215,216,217,221,223,224,225,226,227,229\}\ .
} 
All SG that contains glide(s) have ``a", ``b", ``c", ``d", ``e", or ``n" in their Hermann–Mauguin notations, except SG 113, resulting in
\eqa{
\label{eq:glide_set}
&\text{The list of all SGs that contain glide(s)}= \\
&\{7,9,13,14,15,26,27,28,29,30,31,32,33,34,36,37,39,40,41,43,45,46,48,49,50,51,52,53,54,55,56,57,58,59,60,61,\\
&62,63,64,66,67,68,70,72,73,74,85,86,88,100,101,102,103,104,105,106,108,109,110,112,113,114,116,117,118,120,\\
&122,124,125,126,127,128,129,130,131,132,133,134,135,136,137,138,140,141,142,158,159,161,163,165,167,184,185,\\
&186,188,190,192,193,194,201,203,205,206,218,219,220,222,223,224,226,227,228,230\}\ .
}
Then, the union of the above lists gives us the list of SGs that have at least two currently confirmed AOS symmetries in the presence of TR symmetry and SOC:
\eqa{
&\{2,3,5,6,7,8,9,10,11,12,13,14,15,16,17,18,20,21,22,23,24,25,26,27,28,29,30,31,32,33,34,35,36,37,38,39,\\
&40,41,42,43,44,45,46,47,48,49,50,51,52,53,54,55,56,57,58,59,60,61,62,63,64,65,66,67,68,69,70,71,72,73,74,\\
&75,77,79,80,81,82,83,84,85,86,87,88,89,90,91,92,93,94,95,96,97,98,99,100,101,102,103,104,105,106,107,108,\\
&109,110,111,112,113,114,115,116,117,118,119,120,121,122,123,124,125,126,127,128,129,130,131,132,133,134,\\
&135,136,137,138,139,140,141,142,147,148,149,150,151,152,153,154,155,156,157,158,159,160,161,162,163,164,\\
&165,166,167,168,171,172,174,175,176,177,178,179,180,181,182,183,184,185,186,187,188,189,190,191,192,193,\\
&194,195,196,197,199,200,201,202,203,204,205,206,207,208,209,210,211,212,213,214,215,216,217,218,219,220,\\
&221,222,223,224,225,226,227,228,229,230\}\ ,
}
which consists of 217 SGs.

\subsection{Without TR Symmetry or Without SOC}
Without SOC, we have only three currently confirmed AOS symmetries, inversion, mirror, and glide.
Even if SOC exists, $C_2 \TR$ and $C_4 \TR$ cannot exist in the materials in the absence of TR symmetry since we only consider the 230 SGs instead of the whole set of magnetic SGs.
Therefore, if either TR symmetry or SOC is absent, we should only consider inversion, mirror, and glide symmetries as AOS symmetries for the SG counting here. 
Among the union of \eqnref{eq:inversion_set}-\eqref{eq:glide_set}, the following three sets should be excluded for our purpose:
\eqa{
&\text{The list of the SGs that contain inversion but no mirrors and no glides}=\{2,147,148\}\ ,
}
\eqa{
&\text{The list of the SGs that contain one mirror symmetry but no inversion and no glides}=\{6,8,174\}\ ,
}
and
\eqa{
&\text{The list of the SGs that contain one glide symmetry but no inversion and no mirrors}=\{7,9\}\ .
}
As a result, the remaining 155 SGs consist the list of SGs that have at least two currently confirmed AOS symmetries in the absence of TR symmetry or SOC:
\eqa{
&\{10,11,12,13,14,15,25,26,27,28,29,30,31,32,33,34,35,36,37,38,39,40,41,42,43,44,45,46,47,48,49,50,51,52,53,\\
&54,55,56,57,58,59,60,61,62,63,64,65,66,67,68,69,70,71,72,73,74,83,84,85,86,87,88,99,100,101,102,103,104,\\
&105,106,107,108,109,110,111,112,113,114,115,116,117,118,119,120,121,122,123,124,125,126,127,128,129,130,131,\\
&132,133,134,135,136,137,138,139,140,141,142,156,157,158,159,160,161,162,163,164,165,166,167,175,176,183,184,\\
&185,186,187,188,189,190,191,192,193,194,200,201,202,203,204,205,206,215,216,217,218,219,220,221,222,223,224,\\
&225,226,227,228,229,230\}\ .
}
When TR symmetry is absent, only the following 14 out of 155 SGs allow QAH insulators:
\eqa{
\{10,11,12,13,14,15,83,84,85,86,87,88,175,176\}\ ,
}
while the other 141 SGs forbid QAH insulators, including
\eqa{
&\{25,26,27,28,29,30,31,32,33,34,35,36,37,38,39,40,41,42,43,44,45,46,47,48,49,50,51,52,53,54,55,56,57,58,59,\\
&60,61,62,63,64,65,66,67,68,69,70,71,72,73,74,99,100,101,102,103,104,105,106,107,108,109,110,111,112,113,114,\\
&115,116,117,118,119,120,121,122,123,124,125,126,127,128,129,130,131,132,133,134,135,136,137,138,139,140,141,\\
&142,156,157,158,159,160,161,162,163,164,165,166,167,183,184,185,186,187,188,189,190,191,192,193,194,200,201,\\
&202,203,204,205,206,215,216,217,218,219,220,221,222,223,224,225,226,227,228,229,230\}\ .
}
This counting is simplified by the fact that if a SG $G_0$ forbids QAH insulators, any SG that contains $G_0$ as a subgroup also forbids QAH insulators.
It is because the Hall conductivity is a physical property that is independent of the choices of the unit cell and extra symmetries can only put more constraints on CNs instead of turning a zero CN to nonzero.

\section{More Details on SG 26 with SOC}
\label{app:SG26}

\begin{figure}[t]
    \centering
    \includegraphics[width=0.6\columnwidth]{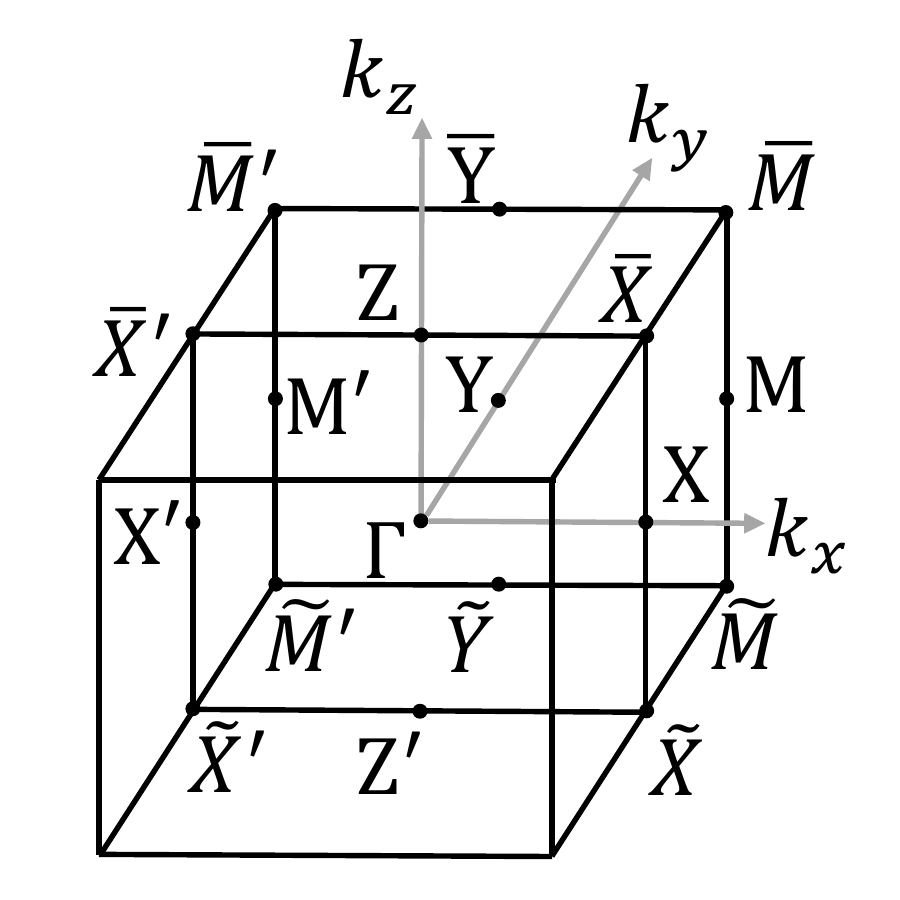}
    \caption{The 1BZ with more labels for SG 26 and SG 13.}
    \label{fig:1BZapp}
\end{figure}

In this section, we discuss SG 26 in more detail.
We first discuss its symmetry representations at high-symmetry momenta, then derive \eqmxgyCN in the main text, and finally discuss the toy model.
The high-symmetry points for this SG include the mirror invariant planes $(k_x=\Lambda)$, the glide invariant planes $(k_y=\Lambda)$, and their intersections, as shown in \figSGtwosix(a) of the main text.
We always assume the system is gapped at all high-symmetry points.
Useful points in 1BZ are labeled according to \figref{fig:1BZapp}.

\subsection{Symmetry Representations}
In this part, we show that the symmetry representations at high symmetry points are the same as atomic insulators.

Owing to the existence of SOC, we have $m_x^2=-1$, $g_y^2=-\{ \mathds{1} | 001\}$, and $\{ m_x,g_y\}=0$.
Without loss of generality, we can always choose the Bloch states to be periodic along $k_x$ and $k_y$, \ie,
\eq{
\ket{\psi_{-\pi,k_y,k_z}}=\ket{\psi_{\pi,k_y,k_z}}\ ,\ \ket{\psi_{k_x,-\pi,k_z}}=\ket{\psi_{k_x,\pi,k_z}}\ .
}

For $k_x=\Lambda$, the occupied Bloch states have definite mirror eigenvalue $\beta\ii$ with $\beta=\pm$, labeled as $\ket{\psi^{m_x,\beta,\alpha}_{(\Lambda,k_y,k_z)}}$.
Here $\alpha=1,...,n^{m_x,\beta}_{(\Lambda,k_y,k_z)}$, and $n^{m_x,\beta}_{(\Lambda,k_y,k_z)}$ is the number of occupied states with mirror eigenvalue $\beta \ii$ at $(\Lambda,k_y,k_z)$.
Similarly, for $k_y=\Lambda$, the occupied Bloch states have definite glide eigenvalue $\beta\ii e^{-\ii k_z/2}$, labeled as $\ket{\psi^{g_y,\beta,\alpha}_{(k_x,\Lambda,k_z)}}$, where $\alpha=1,...,n^{g_y,\beta}_{(k_x,\Lambda,k_z)}$, and $n^{g_y,\beta}_{(k_x,\Lambda,k_z)}$ is the number of occupied states with glide eigenvalue $\beta \ii e^{-\ii k_z/2}$ at $(k_x,\Lambda,k_z)$. 
Since the system is gapped at all high symmetry momenta, $n^{m_x,\beta}_{(\Lambda,k_y,k_z)}$ stay invariant as $(k_y,k_z)$ changes, and so does $n^{g_y,\beta}_{(k_x,\Lambda,k_z)}$ as $(k_x,k_z)$ varies.
Therefore, we may relabel $n^{m_x,\beta}_{(\Lambda,k_y,k_z)}$ and $n^{g_y,\beta}_{(k_x,\Lambda,k_z)}$ as $n^{m_x,\beta}_{k_x=\Lambda}$ and $n^{g_y,\beta}_{k_y=\Lambda}$, respectively.
Along $k_z$, $\ket{\psi^{m_x,\beta,\alpha}_{(\Lambda,k_y,k_z)}}$ can be chosen to be periodic
\eq{
\ket{\psi^{m_x,\beta,\alpha}_{(\Lambda,k_y,-\pi)}}=
\ket{\psi^{m_x,\beta,\alpha}_{(\Lambda,k_y,\pi)}}\ .
}
On the other hand, as $\ket{\psi^{g_y,\beta,\alpha}_{(k_x,\Lambda,\pm \pi)}}$ have opposite glide eigenvalues, we may choose
\eq{
\ket{\psi^{g_y,\beta,\alpha}_{(k_x,\Lambda, -\pi)}}=\ket{\psi^{g_y,-\beta,\alpha}_{(k_x,\Lambda, \pi)}}\ ,
}
which indicates $n^{g_y,-}_{k_y=\Lambda}=n^{g_y,+}_{k_y=\Lambda}$.

At the intersection of the mirror and glide planes $(k_x,k_y)=(\Lambda,\Lambda')$ with $\Lambda'=0,\pi$ , the Hamiltonian is invariant under both mirror and glide operations. 
$\{ m_x,g_y\}=0$ means that the irreducible representations (irreps) here are two-dimensional (2D), where $m_x$ and $g_y$ can be represented as~\cite{BilbaoDoubleGroup}
\eq{
m_x\dot{=}-\ii \mat{1 & \\ & -1}\ ,\  g_y\dot{=}\mat{ & -e^{-\ii k_z}\\ 1 & }\ .
}
We label such 2D irreps as $\Gamma_2$, and label the number of $\Gamma_2$ irreps furnished by occupied states at $(k_x,k_y)=(\Lambda,\Lambda')$ as $n_{(\Lambda,\Lambda')}^{\Gamma_2}$, which is independent of $k_z$ since the system is gapped at the intersection.

Now we include the rest of compatibility relations given by the fact that all high symmetry momenta are fully connected.
First of all, the number of occupied bands should be the same at any high symmetry momentum, labeled as $N$.
Then, one $\Gamma_2$ irrep at $(k_x,k_y)=(\Lambda,\Lambda')$ would split into two bands with opposite mirror (glide) eigenvalues as $k_y$ ($k_x$) move away from the axis. 
Therefore, we have 
\eq{
\frac{N}{2}=n_{(0,0)}^{\Gamma_2}=n_{k_x=0}^{m_x,\pm}=n_{(0,\pi)}^{\Gamma_2}=n_{k_y=\pi}^{g_y,\pm}=n_{(\pi,\pi)}^{\Gamma_2}=n_{k_x=\pi}^{m_x,\pm}= n_{(\pi,0)}^{\Gamma_2}=n_{k_y=0}^{g_y,\pm}\ ,
}
meaning that there are even number of the occupied bands, and $N/2$ is the only independent number.

If we put $N/2$ local orbitals with mirror eigenvalue $+\ii$ at the origin and symmetrize them with SG 26 (2a Wyckoff positions), we get the same symmetry representations as the above.
Therefore, counting symmetry representations cannot identify any gapless phases in this case.

\subsection{Derivation of \eqmxgyCN in the main text}

In this part, we derive \eqmxgyCN in the main text.
For convenience of the discussion, we use $C$ to label the integral of Berry curvature divided by $2\pi$ for certain bands over certain region, $\gamma$ to label the Berry phase divided by $2\pi$ for certain bands along certain path.

As described in the last part, the number $N$ of occupied bands at high symmetry momenta is even, and we still adopt the same boundary condition, \ie, 
\eq{
\label{eq:BC_mx}
 \ket{\psi^{m_x,\beta,\alpha}_{-\pi,k_y,k_z}}=\ket{\psi^{m_x,\beta,\alpha}_{\pi,k_y,k_z}}\ ,\ \ket{\psi^{m_x,\beta,\alpha}_{\Lambda,-\pi,k_z}}=\ket{\psi^{m_x,\beta,\alpha}_{\Lambda,-\pi,k_z}}\ ,\ket{\psi^{m_x,\beta,\alpha}_{\Lambda,k_y,-\pi}}=\ket{\psi^{m_x,\beta,\alpha}_{\Lambda,k_y,\pi}}\ ,
 }
 \eq{
\label{eq:BC_gy}
\ket{\psi^{g_y,\beta,\alpha}_{-\pi,\Lambda,k_z}}=\ket{\psi^{g_y,\beta,\alpha}_{\pi,\Lambda,k_z}}\ ,\ket{\psi^{g_y,\beta,\alpha}_{k_x,-\pi,k_z}}=\ket{\psi^{g_y,\beta,\alpha}_{k_x,\pi,k_z}}\ ,\ \ket{\psi^{g_y,\beta,\alpha}_{k_x,\Lambda,-\pi}}=\ket{\psi^{g_y,-\beta,\alpha}_{k_x,\Lambda,\pi}}\ ,
}
where $\alpha=1,...,N/2$.
The boundary conditions along $k_x$ and $k_y$ are natural, while the boundary condition for $g_y$ eigenstates along $k_z$ is special.
Making such special choice along $k_z$ does not influence the generality of the derivation since both \eqthetamx and \eqnugy in the main text are gauge invariant and independent of the boundary conditions along $k_z$.

Owing to $\left\{ m_x, g_y\right\}=0$, we have 
\eq{
\label{eq:mx_gy_rep}
g_y\ket{\psi^{m_x,\beta,\alpha}_{\Lambda,k_y,k_z}}= \ket{\psi^{m_x,-\beta,\alpha'}_{\Lambda,-k_y,k_z}} V^{g_y,-\beta,\beta}_{\alpha'\alpha}(\Lambda,k_y,k_z)\ ,\ m_x\ket{\psi^{g_y,\beta,\alpha}_{k_x,\Lambda,k_z}}= \ket{\psi^{g_y,-\beta,\alpha'}_{-k_x,\Lambda,k_z}}V^{m_x,-\beta,\beta}_{\alpha'\alpha}(k_x,\Lambda,k_z)\ ,
}
where those $V$ matrices are unitary.
%
Using the above relations and the fact that $C_{\mathscr{A}}-\gamma_{\partial \mathscr{A}}\in \mathds{Z}$ with $\partial \mathscr{A}$ the boundary of $\mathscr{A}$, we can transform \eqthetamx and \eqnugy in the main text to
\eqa{
\nu_{m_x} = ~ & C_{ZZ'\widetilde{Y}\overline{Y}}-C_{\overline{X}\widetilde{X}\widetilde{M}\overline{M}}-2 \gamma^{m_x,+}_{ZZ'}-2 \gamma^{m_x,+}_{\widetilde{Y}\overline{Y}} +2 \gamma^{m_x,+}_{\overline{X}\widetilde{X}}+2 \gamma^{m_x,+}_{\widetilde{M}\overline{M}}\mod 2\ ,\\
\nu_{g_y} = ~ & C_{ZZ'\overline{X}\widetilde{X}}-C_{\overline{M}\widetilde{M}\widetilde{Y}\overline{Y}}-2 \gamma^{g_y,+}_{Z'Z\overline{X}\widetilde{X}Z'}-2 \gamma^{g_y,+}_{\overline{M}\widetilde{M}\widetilde{Y}\overline{Y}\overline{M}}  -2 (\gamma^{g_y,-}_{\overline{X}'\overline{X}}+\gamma^{g_y,-}_{\overline{M}\overline{M}'})\mod 2\ .
}
Adding the two expression together yields
\eqa{
\label{eq:mx_gy_Cs_med}
&\nu_{m_x}+\nu_{g_y}\mod 2\\
&= C_{\mathscr{S}}- 2 \gamma^{+}_{ZZ'}-2 \gamma^{+}_{\widetilde{Y}\overline{Y}} +2 \gamma^{+}_{\overline{X}\widetilde{X}}+2 \gamma^{+}_{\widetilde{M}\overline{M}}-2 \gamma^{g_y,+}_{Z\overline{X}}-2\gamma^{g_y,-}_{\overline{X}'Z}+2\gamma^{g_y,+}_{\overline{Y}\overline{M}}-2\gamma^{g_y,-}_{\overline{Y}\overline{M}'}\mod 2\ ,
}
where 
\eq{
\gamma^{+}_{L}=\gamma^{m_x,+}_{L}-\gamma^{g_y,+}_{L}\ ,
}
and $L$ labels the four lines at the interactions of mirror and glide invariant planes.

What we need to do next is just to show all terms except $C_{\mathscr{S}}$ on the right-hand side of \eqnref{eq:mx_gy_Cs_med} add up to an even number.
First, we can use  \eqnref{eq:mx_gy_rep} to get 
\eqa{
\label{eq:gy_phi}
&-2 \gamma^{g_y,+}_{Z\overline{X}}-2\gamma^{g_y,-}_{\overline{X}'Z}+2\gamma^{g_y,+}_{\overline{Y}\overline{M}}-2\gamma^{g_y,-}_{\overline{Y}\overline{M}'}\mod 2 = -2 \Phi_{V^{m_x,+,-},\overline{X}'Z}-2 \Phi_{V^{m_x,+,-},\overline{Y}\overline{M}'}\mod 2
}
where 
\eq{
\Phi_{U,L}=\frac{-\ii}{2 \pi} \int_L d\bsl{k}\cdot \Tr[U^\dagger(\bsl{k})\nabla_{\bsl{k}}U(\bsl{k})]\ .
}
Now, let us consider $\gamma_{ZZ'}^+$.
With $g_y^2\ket{\psi_{\bsl{k}}}=-e^{-i k_z}\ket{\psi_{\bsl{k}}}$ and $g_y$ being unitary, $V^{g_y}(0,0,k)$ must have the following form 
\eq{
V^{g_y,\beta',\beta}_{\alpha'\alpha}(0,0,k)=
\mat{
0 & W_1(0,0,k)\\
-e^{-\ii k} W^\dagger_1(0,0,k) & 0
}_{\beta'\alpha',\beta\alpha}\ ,
}
where $W_1(0,0,k)$ takes the indexes $\alpha'$ and $\alpha$, is unitary, and satisfies $W_1(0,0,-\pi)=W_1(0,0,\pi)$.
$V^{g_y}_{(0,0,k)}$ can be diagonalized by 
\eq{
U^{g_y,m_x}(0,0,k)=\frac{1}{\sqrt{2}}\mat{
-\ii e^{\ii k/2}W_1(0,0,k) & \ii e^{\ii k/2}W_1(0,0,k)\\
1 & 1
}\mat{
W_2(0,0,k) & 0\\
0 & W_3(0,0,k)
}\ ,
}
where the boundary condition requires $W_2(0,0,\pm \pi)=W_3(0,0,\mp \pi)$ and both $W_2$ and $W_3$ matrices are unitary.
As a result, $V^{m_x}(0,0,k)$ can be expressed by the $W_2$ and $W_3$ as 
\eq{
V^{m_x}(0,0,k)=-\ii 
\mat{
0 & W_2^\dagger(0,0,k)W_3(0,0,k)\\
 W_3^\dagger(0,0,k)W_2(0,0,k) & 0
}\ .
}
With these relations, we can get 
\eq{
2\gamma^+_{Z'Z}\mod 2=-\frac{N}{2}-2 \Phi_{W_2,Z'Z}\mod 2=-\frac{N}{4}+2 \phi_{V^{m_x,+,-},Z}\mod 2\ ,
}
where $\phi_{U,\bsl{k}}=\frac{(-\ii)}{2\pi}\log\det\left[U(\bsl{k})\right]$.
The same derivation can be applied to $\widetilde{Y}\overline{Y}$, $\overline{X}\widetilde{X}$,  and $\widetilde{M}\overline{M}$, resulting in 
\eqa{
& - 2 \gamma^{+}_{ZZ'}-2 \gamma^{+}_{\widetilde{Y}\overline{Y}} +2 \gamma^{+}_{\overline{X}\widetilde{X}}+2 \gamma^{+}_{\widetilde{M}\overline{M}}\mod 2 =2 [\phi_{V^{m_x,+,-},Z}-\phi_{V^{m_x,+,-},\overline{Y}}-\phi_{V^{m_x,+,-},\overline{X}}+\phi_{V^{m_x,+,-},\overline{M}}] \mod 2\\
&= -2 \Phi_{V^{m_x,+,-},Z\overline{X}}-2 \Phi_{V^{m_x,+,-},\overline{M}\overline{Y}}\mod 2\ .
}
Combined the above equation, \eqnref{eq:gy_phi}, \eqnref{eq:mx_gy_Cs_med}, and $\Phi_{V^{m_x,+,-},\overline{X}'\overline{X}/\overline{M}'\overline{M}}\in \mathds{Z}$, we arrive at \eqmxgyCN in the main text.

\subsection{More details on the Tight-Binding Model for SG 26}

In this section, we discuss the toy tight-binding (TB) model for SG 26 in more detail.
There are two sublattice sites in one unit cell, $\bsl{\tau}_1=(0,0,0)$ and $\bsl{\tau}_2=(0,0,1/2)$.
Then, in the real space, the bases of the Hamiltonian read $\ket{\bsl{R}+\bsl{\tau}_i,s}$, and the Fourier transformation of them gives
\eq{
\ket{\bsl{k},\bsl{\tau}_i,s}=\frac{1}{\sqrt{N_l}}\sum_{\bsl{R}} e^{\ii (\bsl{R}+\bsl{\tau}_i)\cdot \bsl{k}} \ket{\bsl{R}+\bsl{\tau}_i,s}\ ,
}
where $s=\uparrow,\downarrow$ the spin index.
The bases clearly satisfy $\ket{\bsl{k}+\bsl{G},\bsl{\tau}_i,s}=e^{\ii \bsl{G}\cdot\bsl{\tau}_i}\ket{\bsl{k},\bsl{\tau}_i,s}$ for any reciprocal lattice vector $\bsl{G}$.
The glide and mirror symmetries are then represented as $g_y\dot{=}-\ii e^{-\ii k_z/2}\tau_x  \sigma_y$ and $m_x\dot{=}-\ii \tau_0  \sigma_x$, where $\tau$'s and $\sigma$'s are Pauli matrices for sublattice and spin indexes.

Based on the symmetry representations, we consider the following Hamiltonian for the toy model
\eq{
H=\sum_{\bsl{k},i,i',s,s'}\ket{\bsl{k},\bsl{\tau}_i,s}[h(\bsl{k})]_{ii',ss'}\bra{\bsl{k},\bsl{\tau}_{i'},s'}\ ,
}
\eq{
\label{eq:h_SG26}
h(\bsl{k})=d_1\tau_z\sigma_x+d_2\tau_y\sigma_0+d_3\tau_x\sigma_0+d_4\tau_z\sigma_z+d_5\tau_y\sigma_x\ ,
}
where 
\eqa{
& d_1=m_0-3+\cos(k_x)+\cos(k_y)+\cos(k_z)\ ,\\
& d_2=\cos(\frac{k_z}{2})\sin(k_y)\ ,\  d_3=\sin(\frac{k_z}{2})\ ,\\
& d_4=\sin(k_x)\ ,\ d_5=m_1 \cos(\frac{k_z}{2})\cos(k_x)\ ,
}
and the $\bsl{k}$ dependence of $d_i$'s is implied.
The eigenvalues of \eqnref{eq:h_SG26} can be analytically solved as 
\eq{
\pm \sqrt{d_1^2+d_3^2+(\sqrt{d_2^2+d_4^2}\pm d_5)^2}\ .
}
We always consider the system at half filling.

From the eigenvalues of $h(\bsl{k})$, we can derive the gapless condition as $d_1=d_3=0$ and $|d_5|=\sqrt{d_2^2+d_4^2}$, which, combined with the expressions of $d_i$'s, gives 
\eqa{
& \cos(k_x)=\frac{-(m_0-2)\pm \sqrt{(m_0-2)^2-(m_1^2+2)((2-m_0)^2-2)}}{m_1^2+2}\ ,\\
& \cos(k_y)=2-m_0-\cos(k_x)\ ,\ k_z=0\ ,\\
& \cos(k_x)\in \mathds{R}\ ,\ |\cos(k_x)|\leq 1\ ,\\
& \cos(k_y)\in \mathds{R}\ ,\ |\cos(k_y)|\leq 1\ .
}
Especially, the gapless points exist on $k_x=0$ plane when 
\eq{
m_0\in [0,2]\ ,\ m_1=\pm \sqrt{m_0(2-m_0)}\ ;
} 
the gapless points exist on $k_y=0$ plane when 
\eq{
m_0\in [0,2]\ ,\ m_1=\pm \sqrt{\frac{1}{(1-m_0)^2}-1}\ .
}
With the above conditions, we plot \figSGtwosix(c) in the main text.

The Fermi arcs shown in \figSGtwosix(d) of the main text are parts of the surface states on the (001) surface.
The Fermi arcs are the intersection between the surface energy dispersion and the $E=0$ plane, as shown in \figSGtwosix(e) of the main text.
As mentioned in the main text, the surface energy dispersion turns out to be in a saddle-surface shape.
It is confirmed by that the energy dispersion bends down along $k_x$-axis but bends up along $k_y$ axis as shown in  \figref{fig:SG26_app} (a) and (b), respectively.
Moreover, we can clearly see that away from the Weyl points, the surface mode typically exists in between the direct bulk gap.

\begin{figure}[t]
    \centering
    \includegraphics[width=0.7\columnwidth]{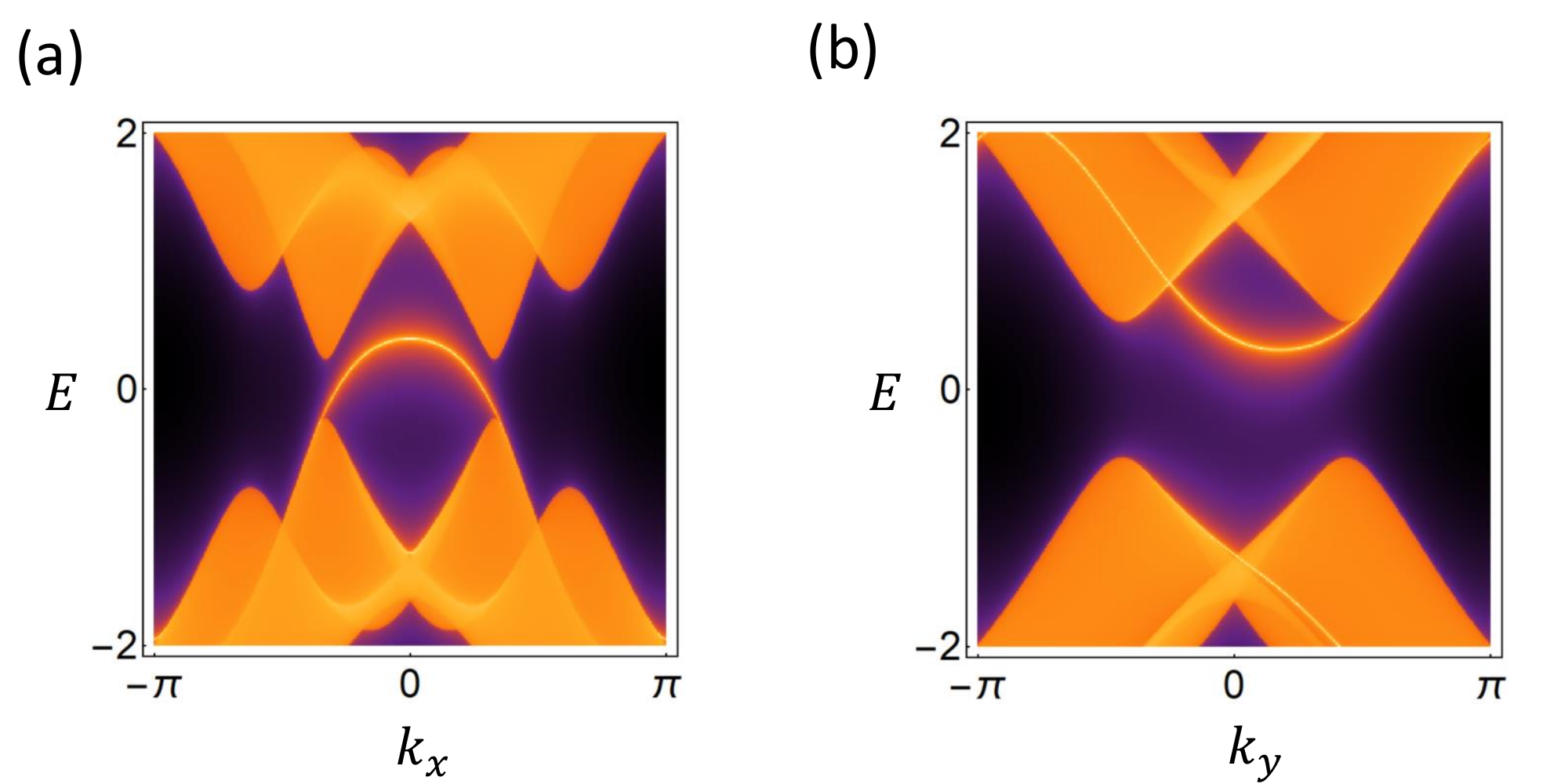}
    \caption{These two figures show the energy dispersion on (001) surface. (a) and (b) are for $k_y=0$ and $k_x=0$, respectively.}
    \label{fig:SG26_app}
\end{figure}

\section{More Details on SG 13 without SOC}
\label{app:P_gy}

In this section, we derive \eqPgymod and \eqPgymodins in the main text for a magnetic crystal with magnetic SG being SG 13 and without SOC in more detail.

\subsection{Derivation of \eqPgymod in the main text}

Recall that we assume that the system is gapped on $k_x=\pi$, $k_y=0$, $k_y=\pi$, and $k_z=\pi$ planes, and has zero CN on $k_y=\pi$.
All the four gapped planes are connected, and we can label the number of occupied band as $N$.
This derivation has overlap with \refcite{Kim2019Glide}.

We first discuss $\nu_{g_y}$.
Since $C_{\mathscr{B}}$ is zero owing to the combination of inversion and glide symmetries $R_2=g_y P$, we only need the consider the remaining parts that are only defined on the $k_y=0,\pi$ planes.
As discussed in the last section, the bands can have definite $g_y$ eigenvalues, $\pm e^{-\ii k_z/2}$, on the $k_y=\Lambda$ plane, and the number of occupied bands with either eigenvalue should be equal to $N/2$, making $N$ a even number.
Then, we can choose the energy eigenstates on $k_y=\Lambda$ as \eqnref{eq:BC_gy} except that $\beta$ now means the $g_y$ eigenvalue $\beta e^{-\ii k_z/2}$ owing to the absence of SOC.
From $P g_y=e^{\ii p_z}g_y P$ and $P^2=1$, we have 
\eq{
\label{eq:V_p}
P\ket{\psi^{g_y,\beta,\alpha}_{k_x,\Lambda,k_z}}=\ket{\psi^{g_y,\beta,\alpha'}_{-k_x,\Lambda,-k_z}} \left[V^{\beta,\beta}_P(k_x,\Lambda,k_z)\right]_{\alpha'\alpha}\ ,
}
and the unitary $V^{\beta,\beta}_P(k_x,\Lambda,k_z)$ satisfies
\eq{
V^{\beta,\beta}_P(k_x,\Lambda,k_z)V^{\beta,\beta}_P(-k_x,\Lambda,-k_z)=1\ .
}
The above relations give~\cite{Turner2012NTRInv,Hughes2011NTRInver}
\eq{
C^{g_y,-}_{k_y=0} \mod 2=2 \gamma^{g_y,-}_{XX'}+2\gamma^{g_y,-}_{\overline{X}'\overline{X}}\mod 2\ ,\ C^{g_y,-}_{k_y=\pi} \mod 2=2 \gamma^{g_y,-}_{MM'}+2\gamma^{g_y,-}_{\overline{M}'\overline{M}}\mod 2\ ,
}
resulting in 
\eq{
\nu_{g_y}=2\gamma_{XX'}^{g_y,-}-2\gamma_{MM'}^{g_y,-}\mod 2\ .
}
Combined with \eqnref{eq:V_p}, we have 
\eq{
\nu_{g_y}=2(\phi_{V^{-.-}_{P},X}-\phi_{V^{-.-}_{P},\Gamma})-2(\phi_{V^{-.-}_{P},M}-\phi_{V^{-.-}_{P},Y})\mod 2\ .
}

All four points, $\Gamma$, $X$, $Y$, and $M$, are on the $k_z=0$ plane.
On $k_z=0$ plane, we have $[P, g_y]=0$ meaning that the states at a inversion-invariant momentum $K_0$ on $k_z=0$ plane can have definite $P$ and $g_y$ eigenvalues simultaneously, where $P$ eigenvalue takes values $\pm 1$.
We label the number of occupied states at $K_0$ with $g_y$ eigenvalue $\beta e^{-\ii k_z/2}$ and $P$ eigenvalue $\beta'$ as $n_{K_0}^{g_y,\beta,P,\beta'}$, and then we have
\eq{
\phi_{V^{-.-}_{P},K_0} \mod 1=\frac{n_{K_0}^{g_y,-,P,-}}{2} \mod 1\ ,
}
resulting in 
\eq{
\nu_{g_y}=\sum_{K_0} n_{K_0}^{g_y,-,P,-} \mod 2\ .
}

Now we simplify the expression of $\nu_P$.
Since $\{P, g_y\}=0$ on $k_z=\pi$, the number of states with $P$ eigenvalue $+1$ is equal to the number of states with $P$ eigenvalue $-1$ at an inversion-invariant momentum on $k_z=\pi$ plane, which further equals to $N/2$.
Combined with that $N$ is even, we can first simply $\nu_P$ to
\eq{
\nu_P=-\sum_{K_0}\frac{n_{K_0}^{P,-}}{2}\mod 2\ .
}
To simply the above equation, first note that $n_{K_0}^{P,-}=n_{K_0}^{g_y,+,P,-}+n_{K_0}^{g_y,-,P,-}$ and $n_{K_0}^{g_y,+,P,-}+n_{K_0}^{g_y,+,P,+}=N/2$.
Since the states at $K_0$ also have definite $R_2$ eigenvalues $\pm 1$, we have $n_{K_0}^{g_y,+,P,+}+n_{K_0}^{g_y,-,P,-}=n^{R_2,+}_{K_0}$ with $n^{R_2,+}_{K_0}$ the number of occupied states with $R_2$ eigenvalue $+1$ at $K_0$.
As a result, we have
\eq{
n_{K_0}^{P,-}=\frac{N}{2}-n_{K_0}^{R_2,+}+2 n_{K_0}^{g_y,-,P,-}\ ,
}
which further results in 
\eq{
\label{eq:P_gy_gen}
\nu_p=\nu_{g_y}+\sum_{K_0} \frac{n_{K_0}^{R_2,+}}{2}\mod 2.
}

To obtain \eqPgymod, the only thing left is to show that 
\eq{
\sum_{K_0} \frac{n_{K_0}^{R_2,+}}{2}-\frac{\Delta_{R_2}}{2}
}
is even, where $\Delta_{R_2}= n_{\Gamma}^{R_2,+}-n_{Y}^{R_2,+}$.
First note that $n^{R_2,+}_X=n^{R_2,+}_M$ since $k_x=\pi$ is gapped, we have 
\eq{
\sum_{K_0} \frac{n_{K_0}^{R_2,+}}{2}-\frac{\Delta_{R_2}}{2}\mod 2
=n^{R_2,+}_{M}+n^{R_2,+}_{Y}\mod 2\ .
}
Since the system has zero CN on $k_y=\pi$ plane, $n^{P,-}_M+n^{P,-}_Y+n^{P,-}_{\overline{M}}+n^{P,-}_{\overline{Y}}$ is even, and we have $n^{P,-}_M+n^{P,-}_Y$ is even owing to $n^{P,-}_{\overline{M}}+n^{P,-}_{\overline{Y}}=N$.
As a result, 
\eqa{
&n^{R_2,+}_{M}+n^{R_2,+}_{Y}\mod 2=n^{g_y,-,P,-}_{M}+n^{g_y,+,P,+}_{M}+n^{g_y,-,P,-}_{Y}+n^{g_y,+,P,+}_{Y}\mod 2\\
&=n^{g_y,-,P,-}_{M}-n^{g_y,+,P,-}_{M}+n^{g_y,-,P,-}_{Y}-n^{g_y,+,P,-}_{Y}\mod 2=n^{P,-}_M+n^{P,-}_Y\mod 2=0\ ,
}
from which we can get \eqPgymod of the main text.

At last, we would like to discuss a bit more about \eqPgymod of the main text.
Since $\nu_{g_y}$ is always integer valued while $\nu_P$ is not, there are two cases that can make $\nu_P\neq \nu_{g_y}$: (i) $\nu_P$=1/2 or 3/2 indicating that the system is gapless~\cite{Turner2012NTRInv,Hughes2011NTRInver}, and (ii) $\nu_P$ is an integer but different from $\nu_{g_y}$.
$\Delta_{R_2}$ is an odd number in case (i) and is twice an odd number in case (ii), both of which indicate the existence of the gapless points on the $\Gamma-Y$ axis given by the crossing between bands with different $R_2$ eigenvalues.
When $\Delta_{R_2}$ is twice an even number and nonzero, the gapless points on the $\Gamma-Y$ axis still exist but they cannot be detected by the gapless criterion. 

\subsection{Derivation of \eqPgymodins in the main text}

We in this part consider a generic magnetic insulator with magnetic SG being SG 13 and without SOC.
The key difference from the last part is that here we do not set the CN on $k_y=\pi$ to zero.
Nevertheless, \eqnref{eq:P_gy_gen} derived above still holds since it does not rely on the EZCN condition.
Therefore, the key here is still the simplification of $\sum_{K_0} \frac{n_{K_0}^{R_2,+}}{2}$.

The insulating condition tells us $n^{R_2,+}_{\Gamma}=n^{R_2,+}_Y$ and $n^{R_2,+}_{M}=n^{R_2,+}_X$.
Then, with $n^{g_y,+,P,+}_{K_0}+n^{g_y,+,P,-}_{K_0}=N/2$ and $N$ even, we have 
\eq{
\sum_{K_0} \frac{n_{K_0}^{R_2,+}}{2}\mod 2=n^{P,-}_{\Gamma}+n^{P,-}_{X}\ \mod\ 2\ .
}
Furthermore, $n^{P,-}_Z=n^{P,-}_{\overline{X}}=N/2$, we eventually arrive at
\eq{
\sum_{K_0} \frac{n_{K_0}^{R_2,+}}{2}\mod 2=n^{P,-}_{\Gamma}+n^{P,-}_{X}+n^{P,-}_{Z}+n^{P,-}_{\overline{X}}\ \mod\ 2=C_{k_y=0}\mod 2\ .
}
Substitute the above expression in the \eqnref{eq:P_gy_gen}, we can get \eqPgymodins in the main text.

\end{widetext}
\bibliography{bibfile_references}

\end{document}